\documentclass{article}
\usepackage{spconf,amsmath,graphicx,color,url, tabularx}
\usepackage{hyphenat}
\usepackage{multirow}
\usepackage{multicol}
\usepackage{caption}
\DeclareMathOperator{\softmax}{softmax}

\title{SynthTab: Leveraging Synthesized Data for Guitar Tablature Transcription\vspace{-1mm}}
\name{Yongyi Zang*, Yi Zhong*, Frank Cwitkowitz, Zhiyao Duan\vspace{-1mm}
\thanks{*Authors contributed equally. This work is partially supported by National Science Foundation (NSF) grants No. 1846184 and No. 2222129, and synergistic activities funded by NSF grant DGE-1922591.}
}

\address{\vspace{-1mm}Department of Electrical and Computer Engineering, University of Rochester, Rochester, NY, USA \\
\footnotesize\texttt{yzang4@u.rochester.edu, yi.zhong@rutgers.edu, fcwitkow@ur.rochester.edu, zhiyao.duan@rochester.edu}
}

\begin{document}
\ninept
\maketitle
\begin{abstract}
Guitar tablature is a form of music notation widely used among guitarists. It captures not only the musical content of a piece, but also its implementation and ornamentation on the instrument. Guitar Tablature Transcription (GTT) is an important task with broad applications in music education, composition, and entertainment. Existing GTT datasets are quite limited in size and scope, rendering models trained on them prone to overfitting and incapable of generalizing to out-of-domain data. In order to address this issue, we present a methodology for synthesizing large-scale GTT audio using commercial acoustic and electric guitar plugins. We procure SynthTab, a dataset derived from DadaGP, which is a vast and diverse collection of richly annotated symbolic tablature. The proposed synthesis pipeline produces audio which faithfully adheres to the original fingerings and a subset of techniques specified in the tablature, and covers multiple guitars and styles for each track. Experiments show that pre-training a baseline GTT model on SynthTab can improve transcription performance when fine-tuning and testing on an individual dataset. More importantly, cross-dataset experiments show that pre-training significantly mitigates issues with overfitting.
\end{abstract}
\section{Introduction}\label{sec:intro}
% \copyrightnote
Automatic Music Transcription (AMT) is the task of converting music audio into some kind of music notation, with broad applications in music education, search, and analysis~\cite{benetos2018automatic}. Guitar Tablature Transcription (GTT) is an instrument-specific characterization of AMT \cite{wu2020multi}, aiming to transcribe guitar audio into guitar tablature. Beyond musical content, tablature specifies the string for each note and any guitar-specific playing techniques that should be employed. With only minimal musical training needed, tablature is highly intuitive for guitarists of all levels. Consequently, it has become the primary means to represent and communicate information regarding guitar performances for both educational and practical purposes\footnote{As evidenced by the large community surrounding websites such as \url{https://www.ultimate-guitar.com}.}. Due to the additional information specified in tablature notation, GTT poses several unique challenges over standard AMT tasks.

Although the utility and necessity of GTT are duly recognized, its progress has been slower relative to other instrument-specific transcription tasks such as piano transcription. One of the main reasons for this is the lack of large-scale annotated datasets. Some existing datasets, such as MedleyDB \cite{bittner2014medleydb}, include a substantial amount of guitar audio but omit specific string-level annotations. Other datasets that include string-level annotations, such as IDMT-SMT-Guitar \cite{kehling2014automatic}, GuitarSet \cite{quingyang2018guitarset}, or EGDB \cite{chen2022towards}, are limited in size and diversity due to significant labor costs in recording and tablature annotation. This issue of data scarcity makes GTT models trained on such datasets susceptible to overfitting issues and limits the development of novel and more advanced transcription methods. In order to continue progressing GTT, innovative and scalable strategies for significantly expanding guitar audio-tablature datasets are imperative.

In this paper, we propose a methodology for synthesizing guitar audio directly from tablature using virtual instrument software. The tablature used for synthesis intrinsically represents the annotations for the resulting audio, allowing us to produce data perfectly suitable for GTT.
We leverage the symbolic tablature dataset DadaGP~\cite{sarmento2021dadagp} and employ our methodology to realize SynthTab, a large-scale, string-accurate, and timbre-rich GTT dataset. Our work bears resemblance to Slakh~\cite{manilow2019cutting} and AAM~\cite{ostermann2023aam}, datasets which comprise audio synthesized from multi-instrument MIDI tracks. However, our approach differs primarily in that we specifically leverage virtual instrument software with string-level note control, and design our synthesis pipeline accordingly. We also synthesize each track with multiple virtual guitars and playing styles, support a subset of playing techniques specified in DadaGP, and incorporate humanization effects such as varying vibrato levels. SynthTab contains roughly 13,113 hours of audio spanning 20,715 tracks and 23 timbral profiles.

In order to investigate the utility of SynthTab, we conduct cross-dataset experiments using three existing GTT datasets featuring real guitar recordings. Our experiments shows that a baseline model trained individually on any of these datasets tends to exhibit poor generalization with respect to the others. We demonstrate that by pre-training on SynthTab, such overfitting can be mitigated, leading to improvements in both same-dataset and cross-dataset scenarios. As the first large-scale guitar audio dataset with tablature annotations, SynthTab paves the way for training larger and more complex machine learning models for GTT and adjacent tasks. Additionally, our proposed synthesis pipeline\footnote{All code and data is made available at \url{www.synthtab.dev}.} can enable the research community to synthesize custom or even larger and more diverse guitar audio-tablature datasets using arbitrary symbolic tablature.

\section{Related Work}\label{sec:related_work}
Recently, there has been increasing attention on the task of GTT, largely driven by the development of guitar audio-tablature datasets. The \textbf{IDMT-SMT-Guitar} \cite{kehling2014automatic} dataset is an early example featuring recordings of various electric guitars, pickup settings, playing styles, and playing techniques. It is split into subsets, three of which include string-level note annotations and contain isolated notes and chords, twelve short licks, and five short pieces, respectively. \textbf{GuitarSet} \cite{quingyang2018guitarset} consists of 360 improvised short recordings of experienced guitarists playing a single acoustic guitar, along with corresponding tablature annotations, acquired through the use of a hexaphonic pickup. Annotations are obtained by de-bleeding and performing pitch tracking on the isolated audio from each separate string. The \textbf{EGDB} \cite{chen2022towards} dataset extends the methodology of GuitarSet to collect 240 clean direct input (DI) audio signals from an experienced guitarist playing a single electric guitar using a hexaphonic pickup. Various pieces of guitar tablature are precisely played with the assistance of a click track, and an alignment procedure based on onset detection is leveraged to obtain high-resolution note annotations. Additional audio is created by feeding the DI signals into several virtual amplifiers.

While these datasets have been immense contributions to the research community, they are relatively limited, with each amounting to only a few hours of audio. Their size pales in comparison to the datasets available for other AMT tasks, such as MAESTRO \cite{hawthorne2018enabling} for piano transcription or E-GMD~\cite{callender2020improving} for drum transcription, which each contain hundreds of hours of data. They also contain little variation in guitar timbre and therefore inharmonicity, a property known to be crucial for differentiating between strings \cite{hjerrild2019estimation}. Recent work on GTT has largely utilized only GuitarSet \cite{wiggins2019guitar, kim2022note, cwitkowitz2023fretnet}, following the cross-validation paradigm. As such, it is unclear whether these approaches can successfully generalize to out-of-domain data.

In parallel, there has also been work on guitar tablature language modeling \cite{chen2020automatic, sarmento2023gtr}, enabled by recent success in sequence modeling \cite{huang2020pop} for music and the proliferation of large collections of symbolic tablature. The \textbf{DadaGP} \cite{sarmento2021dadagp} dataset is one such collection containing symbolic tablature for 26,181 popular songs in the multi-track GuitarPro\footnote{See \url{https://www.guitar-pro.com} for more information.} format. DadaGP offers diverse and high quality guitar tablature at scale, and has created new opportunities for regularizing GTT models \cite{cwitkowitz2022data} and synthesizing realistic data.

\section{Audio Synthesis Pipeline}\label{sec:synth_pipeline}
In this section, we outline the major steps carried out in order to synthesize symbolic tablature from DadaGP \cite{sarmento2021dadagp} and to create SynthTab.

\subsection{Sourcing Tracks from DadaGP}\label{sec:sourcing:tracks}
Given the time and memory requirements associated with rendering audio, and for additional practical considerations, only a subset of the tablature within the multi-track GuitarPro files of DadaGP is selected for synthesis. We synthesize tracks corresponding to acoustic and electric guitars (MIDI instrument numbers 25-31) with six strings and no tempo changes, following a simple procedure to avoid synthesizing duplicate tracks with different GuitarPro versions.

\subsection{Converting GuitarPro to JAMS}\label{sec:to_jams}
Track data is parsed from the respective GuitarPro files using the \texttt{PyGuitarPro} package~\cite{Abakumov2020pyguitarpro}. In order to more conveniently represent the data for synthesis and as ground-truth, information relevant to timing, notes, and a subset of playing techniques is extracted and stored by string using the JAMS~\cite{humphrey2014jams} format, as in GuitarSet \cite{quingyang2018guitarset}.

\subsection{Converting JAMS to MIDI}\label{sec:to_midi}
JAMS annotations for each track are split by string and converted to MIDI data. Virtual instrument software with string-level note control can then be driven with the MIDI to produce audio for each string. Pitch modulation techniques, including \textit{bends} and \textit{vibrato}, are encoded via standard MIDI control channels. Random perturbations are made to the pitch ceiling of notes with vibrato, and a small amount of vibrato is applied randomly to the remaining notes to simulate human playing. Other supported playing techniques, including \textit{hammer-ons}, \textit{pull-offs},  \textit{slides}, \textit{palm-muting}, and \textit{harmonics}, are encoded by region via keyswitches, dedicated MIDI notes recognized by virtual instruments with pitch outside of their playing range.

\subsection{Rendering \& Mixing Audio}\label{sec:to_audio}
Rendering is automated using the DawDreamer package~\cite{braun2021dawdreamer}. 
With little effort, DawDreamer can be used to load MIDI into standalone VST plugins, render the corresponding audio, and store the entirety of the signal in random-access memory. String-level MIDI is rendered individually to ensure proper string usage. Before mixing the audio signals, fundamental frequency annotations surrounding each note are extracted using the YIN~\cite{de2002yin} algorithm. Mixtures are subsequently created by averaging the string-level audio signals.

\subsection{SynthTab}\label{sec:synthtab}
Using the methodology described above, we acquire ground-truth and synthesize symbolic tablature using the Ample Sound guitar plugin suite\footnote{Available at \url{https://amplesound.net/en/index.asp}.}. Acoustic guitar tracks are synthesized using the \textit{L}, \textit{T}, \textit{M}, and \textit{SJ} acoustic guitar plugins. Electric guitar tracks are synthesized using the \textit{SH}, \textit{LP}, \textit{TC}, \textit{VC}, \textit{PF}, \textit{SC}, and \textit{E} electric guitar plugins. All synthesized audio for electric guitar corresponds to DI. Each track is synthesized multiple times with varying styles for each guitar, including \textit{pick} (P), \textit{finger} (F), and \textit{thumb} (Th) style playing, and \textit{bridge} (B), \textit{neck} (N), and \textit{middle} (Mi) pickup settings. In total, there are 7 timbral variations for acoustic guitar and 16 timbral variations for electric guitar. Tables~\ref{tab:variations} and~\ref{tab:distribution} summarize the timbral variation and distribution of tracks in SynthTab. Further statistics related to DadaGP tracks are reported in \cite{sarmento2021dadagp}. All audio is rendered at 24-bit with a 44,100 Hz sampling rate and encoded as FLAC files. Rendering took roughly a week with a 24-core Mac Studio M2 Ultra.

\begin{table}[t]
\centering
\begin{tabular}{|c|c|c|c|c|c|c|c|c|}
\hline
& \textit{P} & \textit{F} & \textit{Th} & \textit{B} & \textit{N} & \textit{B}/\textit{N} & \textit{B}/\textit{Mi} & \textit{Mi}/\textit{N} \\
\hline
\textit{L} & $\times$ & $\times$ & & & & & & \\
\hline
\textit{T} & $\times$ & $\times$ & & & & & & \\
\hline
\textit{M} & $\times$ & $\times$ & & & & & & \\
\hline
\textit{SJ} & & & $\times$ & & & & & \\
\hline
\hline
\textit{SH} & $\times$ & $\times$ & & & & & & \\
\hline
\textit{TC} & & & & $\times$ & $\times$ & & & \\
\hline
\textit{VC} & & & & $\times$ & $\times$ & & & \\
\hline
\textit{LP} & & & & $\times$ & & $\times$ & & \\
\hline
\textit{PF} & & & & $\times$ & $\times$ & $\times$ & & \\
\hline
\textit{SC} & & & & $\times$ & $\times$ & & $\times$ & $\times$ \\
\hline
\textit{E} & & & & & $\times$ & & & \\
\hline
\end{tabular}
\caption{Timbral variations in SynthTab, organized by guitar plugin (rows) and playing style (columns). See Section~\ref{sec:synthtab} for more details.}
\label{tab:variations}
\end{table}

\begin{table}[t]
\centering
\begin{tabular}{|c|c|c|c|}
\hline
\textbf{Midi Guitar} & \textbf{\# Tracks} & \textbf{Variations} & \textbf{Total Hours} \\
\hline
\textit{Acoustic Nylon} (25) & 5501 & \multirow{2}{*}{7} & 1620.40 \\
\cline{1-2} \cline{4-4}
\textit{Acoustic Steel} (26) & 5149 & & 1890.95 \\
\hline
\hline
\textit{Electric Jazz} (27) & 1572 & \multirow{5}{*}{16} & 1305.73 \\
\cline{1-2} \cline{4-4}
\textit{Electric Clean} (28) & 2989 & & 2793.04 \\
\cline{1-2} \cline{4-4}
\textit{Electric Muted} (29) & 504 & & 467.47 \\
\cline{1-2} \cline{4-4}
\textit{Overdriven} (30) & 1556 & & 1534.21 \\
\cline{1-2} \cline{4-4}
\textit{Distortion} (31) & 3444 & & 3501.09 \\
\hline
\end{tabular}
\caption{SynthTab track distribution by MIDI instrument.}
\label{tab:distribution}
\end{table}

\section{Experiments}\label{sec:experiments}
Since the cues that help differentiate between strings are highly dependent on instrument-specific properties, cross-dataset evaluation is especially important for GTT. We conduct experiments where a baseline model is trained and evaluated on real guitar recordings from the IDMT-SMT-Guitar \cite{kehling2014automatic} (abbreviated IDMT), GuitarSet \cite{quingyang2018guitarset}, and EGDB \cite{chen2022towards} datasets, which each contain unique timbral properties (see Section~\ref{sec:related_work} for more details). To our knowledge, this is the first attempt to perform cross-dataset benchmarking for GTT. For our experiments, we only utilize the twelve-lick subset of IDMT, due to limited content within the other subsets, and the clean DI signals from EGDB, to maintain consistency with the audio rendered for SynthTab. We create training, validation, and testing splits randomly by track for IDMT and GuitarSet, following 10:1:1 and 8:1:1 ratios, respectively. For EGDB we adopt the pre-defined 8:1:1 split.

In order to explore the capacity of SynthTab to progress research on GTT, we then utilize a larger version of the model and perform the same set of experiments with and without pre-training on SynthTab. Due to potential conflicts between tablature distributions, we only train on data originating from tracks corresponding to \textit{Acoustic Nylon}, \textit{Acoustic Steel}, \textit{Electric Jazz}, and \textit{Electric Clean}. We also hold out 10\% of tracks and the \textit{SJ-Th} and \textit{E-N} timbral profiles for validating the model during the pre-training phase.

\subsection{Baseline Model}\label{sec:model}
We adopt TabCNN \cite{wiggins2019guitar}, one of the earliest DNNs proposed for GTT, as the baseline model in order to carry out all experiments. TabCNN is a lightweight convolutional neural network (CNN) which produces string-level fret class estimates for fixed-length windows of the Constant-Q Transform (CQT) \cite{brown1991calculation} of an audio signal. Classes consist of silence, open string, and the first 19 frets, and are predicted separately for each string using $\softmax$ activation. Since there are six strings, the output of TabCNN is a 126-dimensional six-hot vector. For the second set of experiments with and without pre-training, we add more complexity to the model by quadrupling the number of filters in each convolution layer. This version of the model is referred to as TabCNNx4. Although more recent GTT models have been proposed, no models have yet been subject to cross-dataset evaluation, and TabCNN is comparatively simple and yields solid performance.

\subsection{Evaluation Metrics}\label{sec:metrics}
All models are evaluated using the GTT metrics proposed in \cite{wiggins2019guitar}, which measure frame-level tablature and multi-pitch estimation proficiency. However, we only report $F_1$-score \% ($F_1$) for a more compact presentation of results. Fret class predictions must be made on the correct string for tablature estimation, whereas only the nominal pitch of predictions are considered for multi-pitch estimation. Tablature $F_1$ signifies the ability of a model to correctly estimate pitch activity and differentiate between strings, which is essential for GTT. As such, it is used as the validation criterion to select the best model for each experiment. Both $F_1$-scores are averaged across all tracks within the validation or evaluation set for final scores.

\subsection{Training Details}\label{sec:training}
All training and fine-tuning is done with AdaDelta optimizer using an initial learning rate of $1.0$ and batch size 32 on a single NVIDIA RTX 4090. When fine-tuning pre-trained models, the initial learning rate and batch size are reduced to 0.1 and 8, respectively. Prior to training, all audio is downsampled to 22,050 Hz. CQT features with 192 bins, 24 bins per octave, and a hop size of 512 are computed for each track during the feature extraction stage. Batches are created randomly from the respective training datasets, and each track is sampled only once per epoch with a sequence length of 500 frames.

\section{Results}\label{sec:results}

\begin{figure*}[]
\centering
\includegraphics[width=\textwidth]{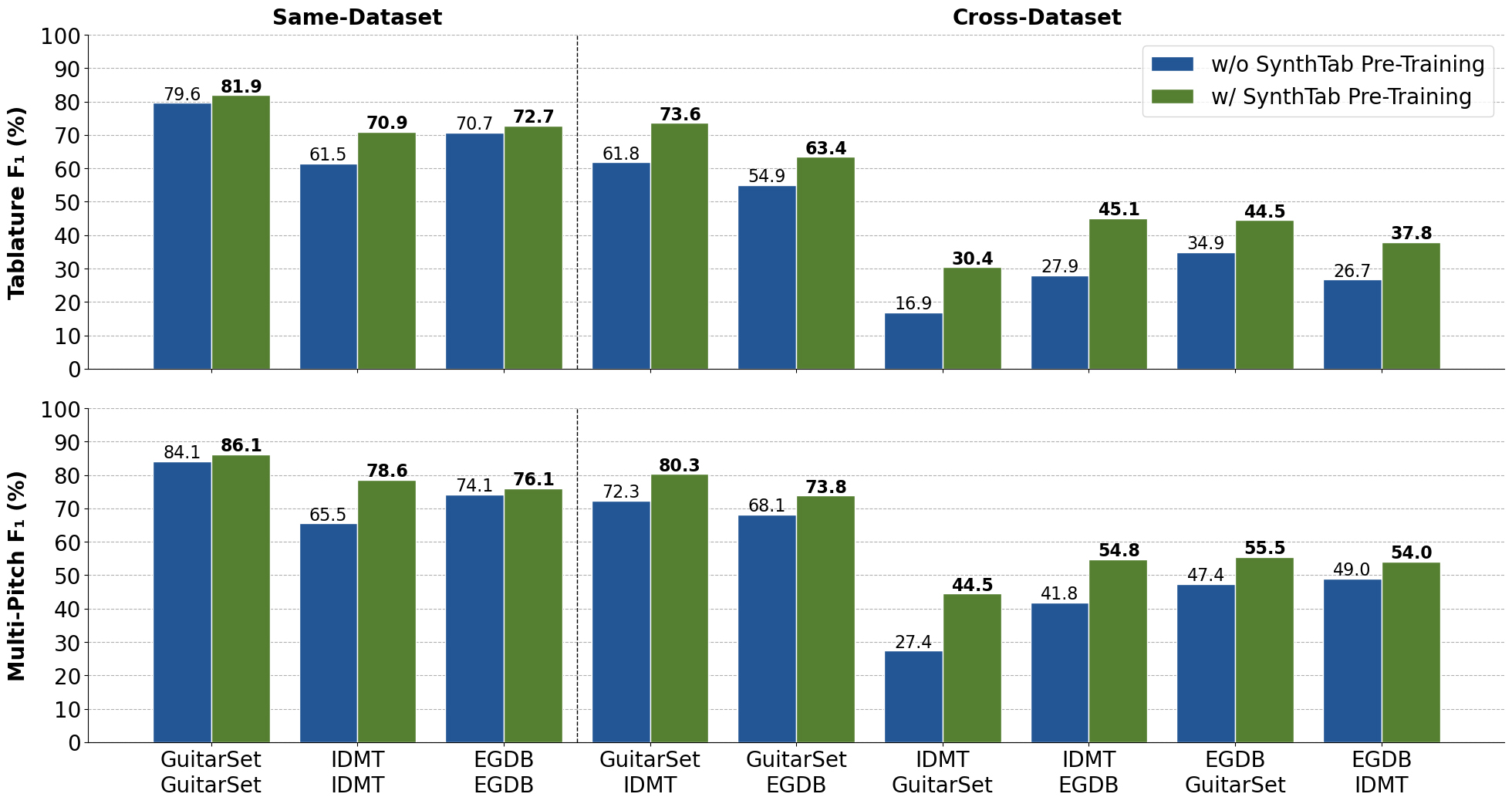}
\caption{Transcription results for TabCNNx4 with and without pre-training on SynthTab when trained or fine-tuned on individual datasets and evaluated on the testing subset of each dataset. Training or fine-tuning datasets are listed above testing datasets for each experiment scenario.}
\label{fig:results}
\end{figure*}

\subsection{Initial Cross-Dataset Benchmarking}\label{sec:results_benchmarking}
The initial cross-dataset benchmarking results for TabCNN are presented in Table~\ref{table:tabcnn_baseline}. As a sanity check, we note that the performance of TabCNN trained and tested on GuitarSet is comparable\footnote{Note that the original paper conducted six-fold cross-validation.} to what was originally reported in \cite{wiggins2019guitar}. However, it is clear that in each case the model has trouble generalizing to unseen data. Performance for the unmatched datasets is substantially weaker than that of the matched dataset for all three experiments. This issue is more prominent for tablature estimation, and is likely caused by the domain mismatch across GuitarSet, IDMT, and EGDB, which each feature different guitars and recording conditions. These experiments indicate that timbral features learned by a GTT model for an individual dataset exhibit low transferability under this experimental setup.

\begin{table}[ht]
\centering
\begin{tabular}{|cc|ccc|}
\hline
\multicolumn{2}{|c|}{\multirow{2}{*}{\textbf{Tablature $F_1$ (\%)}}} & \multicolumn{3}{c|}{\textbf{Test}} \\
\cline{3-5}
\multicolumn{2}{|c|}{} & \multicolumn{1}{c|}{\textbf{GuitarSet}} & \multicolumn{1}{c|}{\textbf{IDMT}} & \textbf{EGDB} \\
\hline
\multicolumn{1}{|c|}{\multirow{3}{*}{\textbf{Train}}} & \textbf{GuitarSet} & \multicolumn{1}{c|}{\textbf{80.1}} & \multicolumn{1}{c|}{57.9} & 55.2 \\
\cline{2-5}
\multicolumn{1}{|c|}{} & \textbf{IDMT} & \multicolumn{1}{c|}{15.3} & \multicolumn{1}{c|}{\textbf{63.1}} & 28.8 \\
\cline{2-5}
\multicolumn{1}{|c|}{} & \textbf{EGDB} & \multicolumn{1}{c|}{37.1} & \multicolumn{1}{c|}{23.5} & \textbf{70.5} \\
\hline
\hline
\multicolumn{2}{|c|}{\multirow{2}{*}{\textbf{Multi-Pitch $F_1$ (\%)}}}  & \multicolumn{3}{c|}{\textbf{Test}} \\
\cline{3-5}
\multicolumn{2}{|c|}{} & \multicolumn{1}{c|}{\textbf{GuitarSet}} & \multicolumn{1}{c|}{\textbf{IDMT}} & \textbf{EGDB} \\
\hline
\multicolumn{1}{|c|}{\multirow{3}{*}{\textbf{Train}}} & \textbf{GuitarSet} & \multicolumn{1}{c|}{\textbf{84.5}} & \multicolumn{1}{c|}{72.5} & 67.2 \\
\cline{2-5}
\multicolumn{1}{|c|}{} & \textbf{IDMT} & \multicolumn{1}{c|}{27.7} & \multicolumn{1}{c|}{\textbf{66.8}} & 43.4 \\
\cline{2-5} 
\multicolumn{1}{|c|}{} & \textbf{EGDB} & \multicolumn{1}{c|}{48.2} & \multicolumn{1}{c|}{45.1} & \textbf{74.0} \\
\hline
\end{tabular}
\caption{Transcription results for TabCNN when trained on individual datasets and evaluated on the testing subset of each dataset.}
\label{table:tabcnn_baseline}
\end{table}

\begin{table}[ht]
\centering
\begin{tabular}{|c|c|c|c|c|}
\hline
\multirow{2}{*}{$F_1$ (\%)} & \textbf{Val.} & \multicolumn{3}{c|}{\textbf{Test}} \\
\cline{2-5}
 & \textbf{SynthTab} & \textbf{GuitarSet} & \textbf{IDMT} & \textbf{EGDB} \\
\hline
\multicolumn{1}{|c|}{\textbf{Tablature}} & 64.2 & 43.1 & 13.8 & 57.0 \\
\hline
\multicolumn{1}{|c|}{\textbf{Multi-Pitch}} & 77.4 & 70.2 & 74.2 & 74.4 \\
\hline
\end{tabular}
\caption{SynthTab validation results and individual dataset testing results for TabCNNx4 pre-trained on SynthTab with no fine-tuning.}
\label{table:synthtab_pretrain}
\end{table}

\subsection{Investigating Pre-Training on SynthTab}\label{sec:results_pretraining}
The results for TabCNNx4 pre-trained on SynthTab, without any further fine-tuning, are presented in Table~\ref{table:synthtab_pretrain}. In general, the pre-trained model appears to follow the same trends as described in Section~\ref{sec:results_benchmarking}. There are several possible explanations for this phenomenon, including once again domain mismatch between SynthTab and the evaluation datasets, training data that is too broadly distributed with no specificity (\textit{e.g.} acoustic-only, electric-only, or data from a specific guitar model), or use of a GTT model with insufficient complexity. We observe that the tablature estimation performance of the pre-trained model improves slightly with respect to the unmatched experiments on GuitarSet and EGDB, though this could be due to the larger model. Multi-pitch estimation performance improves for all unmatched experiments, indicating that SynthTab can provide a solid foundation for training more general transcription models.

The cross-dataset results for TabCNNx4 with randomly initialized weights versus the model pre-trained on SynthTab are plotted in Figure~\ref{fig:results}. Without pre-training, TabCNNx4 achieves similar performance to TabCNN under all training and testing scenarios for both tablature estimation and multi-pitch estimation. In all same-dataset and cross-dataset experiments, pre-training on SynthTab yields improved performance, which in some cases is quite substantial. The relative improvement in most cross-dataset experiments is more pronounced for tablature estimation, meaning it cannot be explained wholly by more robust multi-pitch estimation. In several cross-dataset experiments, fine-tuning lowered performance on the testing dataset relative to the pre-trained model without fine-tuning. However this is not too surprising, given the domain mismatch between the individual datasets and the small amount of fine-tuning data available in some cases. Interestingly, for the pre-trained model fine-tuned on GuitarSet, the performance on IDMT actually exceeds that of the corresponding same-dataset experiments.

\subsection{Discussion}\label{sec:discussion}
Our results show that SynthTab has the potential to benefit GTT models by helping them learn more general timbral features and by improving their underlying multi-pitch estimation robustness. Models pre-trained on SynthTab and fine-tuned on subsequent data unanimously outperform the models that are only trained on the individual datasets. This observation validates the utility of the proposed synthesis methodology and the use of SynthTab for GTT. However, the results also suggest that there are significant challenges in training GTT models to generalize to completely unseen guitar data.

Future work will consist of further investigating the issue of generalization and further exploring the utility of SynthTab for GTT and related tasks. We plan to expand and improve the quality of dataset, with particular focus on better humanization strategies, varying the virtual recording environment and post-processing techniques for synthesis, and the incorporation of data augmentation for more effective training. Finally, with the realization of SynthTab, we will proceed with novel and more advanced model development focused on leveraging the scale and diversity of the dataset.

\section{Conclusion}
In this work, we presented SynthTab, the first large-scale synthesized guitar audio dataset with string-accurate tablature annotations. Utilizing the large collection of symbolic tablature available in DadaGP and commercial virtual instrument software, our proposed synthesis pipeline produces high-quality, string-accurate audio with varied timbre. We have shown through cross-dataset experimentation that training a baseline model on existing guitar audio datasets leads to poor generalization due to domain mismatch and the homogeneity of such datasets. We also demonstrate that pre-training on SynthTab and fine-tuning on individual datasets can lead to more robustness and improved generalization capacity. SynthTab creates new opportunities for advancing guitar transcription research, and future work will consist of further expansion, improvement, and exploration.

\pagebreak
\onecolumn
\begin{multicols}{2}
\bibliographystyle{IEEEbib}
\bibliography{synthtab}
\end{multicols}

\end{document}